\documentstyle[12pt]{article}
\input epsf
\def\be{\begin{equation}}
\def\ee{\end{equation}}
\def\l{\label}
\def\refe#1{(\ref{#1})}

\def\eg{{\it e.g.}}
\def\ie{{\it i.e.}}

\def\LEP2{{LEPII}}
\def\mg{$m_{3/2}$}
\def\th{$\theta$}
\begin{document}
\begin{titlepage}
\vspace*{-1.5cm}
\begin{center}
\null
\vskip-1.5truecm
\rightline{INS-Rep-1189}
\vskip1truecm
\null
\vskip-1.5truecm
\rightline{DPSU-97-4}
\vskip1truecm
\vspace{5ex}
{\Large {\bf  Phenomenological implications of moduli-dominant
SUSY breaking}}\\
\vspace{3ex}
{\bf Yoshiharu Kawamura}$^{a}$, {\bf Shaaban Khalil}$^{b,c}$ and {\bf
Tatsuo
Kobayashi}$^{d}$\\
{\it
\vspace{1ex} a)Department of Physics, Shinshu University,\\
Matsumoto, 390 Japan.}\\
{\it
\vspace{1ex} b) International Center For Theoretical Physics,\\
 ICTP, Trieste, Italy.}\\
{\it
\vspace{1ex} c) Ain Shams University, Faculty of Science\\
 Department of Mathematics, Cairo, Egypt.}\\
{\it
\vspace{1ex} d)Institute for Nuclear Study, University of Tokyo,\\
Midori-cho, Tanashi, Tokyo 188, Japan.}\\
\vspace{6ex}
{ABSTRACT}  
\end{center}
\begin{quotation}
We study moduli-dominated SUSY breaking within the framework of
string models.
This type of SUSY breaking in general leads to non-universal soft masses,
\ie\ soft scalar masses and gaugino masses.
Further gauginos are lighter than sfermions.
This non-universality has phenomenologically important implications.
We investigate radiative electroweak symmetry breaking in the mass spectrum
derived from moduli-dominated SUSY breaking,
where the lightest chargino and neutralino are almost gauginos.
Moreover, constraints from the branching ratio of $b \rightarrow s \gamma$
and the relic abundance of the LSP are also considered.
The mass spectrum of moduli-dominated SUSY breaking is favorable
to the experimental bound of the $b \rightarrow s \gamma$ decay
decreasing its branching ratio.
We obtain an upper bound for the gravitino mass from
the cosmological constraint.
\end{quotation} 
\end{titlepage}
\vfill\eject

\section{Introduction}
\hspace{0.75cm}Supersymmetry (SUSY) is one of the most important keywords 
beyond the standard model \cite{Nilles}.
It is expected that superpartners as well as Higgs particles
would be detected in the near future.
Thus it is very important to study which type of mass spectra for
superpartners and Higgs particles and which type of phenomenological aspects
are predicted from fundamental theory such as grand unified theories
(GUTs) or superstring theory.\\
\vskip -0.3cm
Superstring theory is a promising candidate for unified theory including
gravity. Hence it is interesting to predict the mass spectrum and other
SUSY phenomenological aspects like flavor changing processes,
electric dipole moment and rare decay as well as cosmological aspects
within the framework of superstring theory.
We have not understood SUSY breaking mechanism completely yet.
However we can parameterize unknown $F$-terms and write soft SUSY breaking
terms when we assume the fields that contribute to SUSY breaking
\cite{IL,ST-soft}.
Generic string models include a dilaton field $S$ and moduli fields $T_m$
in their massless spectra.
Vacuum  expectation values (VEVs) of these fields correspond to
the coupling constant and geometrical feature of a six-dimensional
compactified space.
When these fields contribute to SUSY breaking, soft SUSY breaking terms 
can be written in terms of the gravitino mass $m_{3/2}$ and
some goldstino angles under the assumption of the vanishing cosmological
constant \cite{BIM,multiT}.\\
\vskip -0.3cm
Recently there have been various works devoted to study
phenomenological implications of these soft SUSY breaking terms
which are written in terms of the gravitino mass and goldstino angles
only. The dilaton-dominated SUSY breaking leads to universal soft SUSY
breaking terms. Phenomenological implications of this universality were 
studied in Refs.\cite{Sdomi,stringbs}, e.g.  successful electroweak 
symmetry
breaking and the $b \rightarrow s \gamma$ process.
We have non-universal soft SUSY breaking terms when both moduli fields and
the dilaton field contribute to SUSY breaking.
However, such non-universality is not strong.
Thus such a mixed case leads to almost similar phenomenological aspects
of the dilaton-dominated case.
Actually some phenomenological aspects of the mixed case were   
studied in Refs.\cite{Khalil,stringbs}, which include the case of the overall
modulus $T$ with the modular weights chosen such
that one can have appropriate large string threshold corrections to fit
the joining of gauge coupling constants
at GUT scale \cite{BIM}.\footnote{See also Ref.\cite{stringU}.}\\
\vskip -0.3cm
        On the other hand, the moduli-dominated case can lead to strong
non-universality among soft SUSY breaking terms \cite{BIM,multiT}.
The moduli-dominated case with the overall moduli field leads to
non-universal gaugino masses, but a universal soft scalar mass~\cite{BIM}.
Further multi-moduli cases can lead to non-universal soft
scalar masses as well as non-universal gaugino masses \cite{multiT}.   
It is important to study phenomenological aspects of this string inspired
non-universality in order to understand the whole allowed parameter space of
soft SUSY breaking parameters derived from superstring theory.\footnote{
Phenomenological implications of non-universal soft SUSY
breaking terms were discussed in Ref.\cite{nonuni} within the framework of
generic SUSY GUT and supergravity theory, but not 
superstring 
theory.} In Ref.\cite{CDG} non-universal gaugino masses were discussed   
within the overall moduli-dominated case, showing interesting results.\\
\vskip -0.3cm
        In this paper we study the multi-moduli dominated SUSY breaking,
especially phenomenological implications of strong non-universality
among soft scalar masses as well as gaugino masses.
Orbifold models are simple and interesting 4-dimensional string
models \cite{Orbi} and
these models include three diagonal moduli fields $T_m$ ($m=1,2,3$).
Here we discuss phenomenological implications of strong non-universality
led in the case that these three diagonal moduli fields contribute
dominantly to SUSY breaking.
We consider mass spectrum, successful electroweak symmetry breaking,
the $b \rightarrow s \gamma$ process and cosmological
constraints, e.g, the relic abundance of the lightest neutralino. We find
that there appears a significant splitting between the stop masses 
and this leads
to interesting phenomenological implications. Moreover we show that the
lightest stop is the lightest sfermion of this model. \\
\vskip -0.3cm
          
        Study on the $b \rightarrow s \gamma$ process gives a strong constraint
on generic supersymmetric models \cite{bs}, because, in general, there are
additive contributions to the standard model (SM) one. On the other hand,
by including the
next-to-leading order in QCD correction for $b \rightarrow s \gamma$
decay, the SM prediction is above the CLEO measurement at the $1\sigma$
level~\cite{misiak}. Hence, it is likely that SUSY models provide us with
a substantial destructive interference. We show that in moduli-dominated
SUSY breaking the chargino contribution which gives a destructive
interference is dominant and the values of the total branching ratio for
all the parameter space are less than the SM branching ratio. This result
corresponds to the case of $\mu >0$ which gives strong mass splitting
between the stop masses according to our sign convention.\\
\vskip -0.3cm
                We also show that the lightest neutralino is
the lightest superpartner (LSP) and it is found to be dominantly bino. In
contrast with the
case of the overall modulus analyzed in Ref.~\cite{CDG} the LSP mass is not
degenerate with 
the chargino mass. The relic abundance of the LSP imposes important
constraints on the gravitino mass \mg\ and requires a large value of
$\theta_1$\ (one of goldstino angles) \ie\ requires strong non-universality
between the down and up
sector Higgs soft masses.\\
\vskip -0.3cm

        This paper is organized as follows. In section 2, we review
formulae of soft SUSY breaking terms within the framework of string
models.
In addition, we specify our model with typical non-universal soft masses,
which are derived from string models. In section 3, we study the radiative
electroweak breaking and we determine the particle spectrum of the model
for
both cases of $\mu > 0$ and $\mu <0$. Section 4 is devoted to the
constraints on the parameter space due to $b \rightarrow s \gamma$ decay.
We show that in the case of $\mu > 0$ the CLEO measurement does not impose
any constraint on the parameter space. In section 5, we study the relic
abundance of the lightest neutralino and we show that it imposes an upper
bound of order 1.5 TeV on the gravitino mass and requires that $\theta_1$
be large. In section 6, we analyze the effect of changing the
value of the parameter $T$, namely we study the implication of the small
values of $T$ as well as the large value of $T$. Finally we give our
conclusions in section 7.

\section{Moduli-dominated SUSY breaking}
\hspace{0.75cm}In orbifold models with three diagonal moduli fields $T_m$ 
($m=1,2,3$)
the K\"ahler potential is obtained as
\begin{eqnarray}
K= - log (S+S^*) - \sum_{m=1}^3 log(T_m+T_m^*) + \sum_i
\phi_i \phi_i^* \prod_{m=1}^3(T_m+T_m^*)^{n_i^m},
\end{eqnarray}
where $n_i^m$ is the modular weight of the field $\phi_i$ for the
$m$-th moduli field $T_m$ \cite{OrbSG2}.
In addition, the gauge kinetic function $f_a$ is obtained at tree level as
$f_a=k_aS$, where $a$ is an index for the gauge group and $k_a$ is
the corresponding Kac-Moody level.\\
\vskip -0.3cm 
Here we assume that  $S$ and $T_m$ contribute to SUSY breaking with a
nonperturbative superpotential $W(S,T_m)$, although we do not specify
its form.
Then we parameterize unknown $F$-terms as follows \cite{BIM,multiT},
\begin{eqnarray}
(K_{S}^S)^{1/2} F^S = \sqrt{3} m_{3/2} \sin \theta, \quad
(K_{T_m}^{T_m})^{1/2} F^{T_m} = \sqrt{3} m_{3/2} \cos \theta \Theta_m,
\end{eqnarray}
where $\sum_{m=1}^3 \Theta_m^2=1$.
Here gravitino mass $m_{3/2}$ is defined as
$m_{3/2}=\langle e^{K/2}W \rangle $.
Using the above parametrization with 
$$\Theta_1=\sin\theta_1 \sin\theta_2, \hspace{1cm} \Theta_2=\sin\theta_1
\cos\theta_2, \hspace{1cm} \Theta_3=\cos\theta_1$$
we can write the soft scalar mass $m_i$ and the gaugino mass $M_a$ as
\begin{eqnarray}
m_i^2 &=& m_{3/2}^2 (1+3 \cos^2 \theta \sum_m n_i^m \Theta_m^2 ),
\label{softm} \\
M_a &=& \frac{\sqrt{3} m_{3/2}}{Re f_a}\left[k_a Re S \sin \theta
+\cos \theta \sum_m (b'^m_a-k_a \delta_{GS}^m) D(T_m,T_m^*)
\Theta_m \right],
\label{gauginoM}
\end{eqnarray}
where the second term on the right-hand side of eq.(\ref{gauginoM}) is
due to moduli-dependent threshold corrections
at one-loop level \cite{thres,DFKZ}.
Here the function $D(T)$ is given by the use of 
the Eisenstein function $\widehat{G(T)}$
as \footnote{Several kinds of modular functions are shown
in Ref.\cite{CFILQ}.}
\begin{eqnarray}
D(T) = \frac{(T+T^*)}{32\pi^3} \widehat{G(T)}.
\end{eqnarray}  
For example, the values of $D(T)$ are $1.5 \times 10^{-3}$, $2.7 \times
10^{-2}$,
$6.0 \times 10^{-2}$ and $6.6 \times 10^{-1}$ for $T=1.2$, $5.0$, $10$ and
$100$, respectively.
In addition $\delta_{GS}^m$ is the Green-Schwarz coefficient \cite{GS},
which is gauge group-independent,  and
$b'^m_a$ denotes a duality anomaly coefficient given by \cite{DFKZ}
\begin{eqnarray}
b'^m_a = -C(G_a) + \sum_R T(R) (1+2n_a^m),
\end{eqnarray}
where $C(G_a)$ is the casimir of the adjoint representation and
$T(R)$ is the index of the $R$ representation.
Further we obtain the $A$-term associated to the $T$-independent
Yukawa coupling as  
\begin{eqnarray}
A_{ijk}=-\sqrt{3} m_{3/2} \left[ \sin \theta +
\cos\theta \sum_m (1+n_i^m+n_j^m+n_k^m) \Theta_m   \right],
\label{Aterm}   
\end{eqnarray}
where $n_i^m$, $n_j^m$ and $n_k^m$ are modular weights of the fields
to couple.
One needs a correction term in eq.(\ref{Aterm}) when the corresponding
Yukawa coupling depends on moduli fields.
However, $T$-dependent Yukawa coupling in general includes suppression
factors \cite{Yukawa}.
Thus strong Yukawa coupling such as the top Yukawa coupling is
expected to be independent of $T$.
The above formula (\ref{Aterm}) seems to be reasonable.\\
\vskip -0.3cm
Finally, we have to consider the scalar bilinear soft breaking term $B 
\mu H_1 H_2$, i.e., the $B$-term,
where $H_1$ and $H_2$ are the down and up sectors of Higgs fields, 
respectively.
The form of the $B$-term strictly
depends on the origin of the $\mu$-term in the superpotential and/or
the K\"ahler potential.
In Ref.~\cite{BIM} three sources for the $B$ parameter were considered,
labeled by $B_Z$, $B_{\mu}$ and $B_{\lambda}$. The source of $B_Z$ is
the presence of certain bilinear terms in the K\"ahler potential which
can naturally induce a $\mu$-term of order $m_{3/2}$ 
after SUSY breaking \cite{mu1}. An alternative mechanism to
generate a $B$-term in the scalar potential is to assume that the
superpotential $W$ includes a SUSY mass term $\mu(S,T_i) H_1 H_2$ 
induced by a non-perturbative effect, 
then a $B$-term is automatically generated
and it is called $B_{\mu}$.
Also it was pointed out~\cite{mu2} that
the presence of a non-renormalizable term in the superpotential $\lambda
W H_1 H_2$ yields dynamically a $\mu$ parameter when $W$ acquires VEV
$\mu(S,T_i) = \lambda(T_i) W(S, T_i)$ and the corresponding $B$-term is
denoted by $B_{\lambda}$. In general, there could be an admixture of the   
above three possibilities.
Thus we will treat $B$ as a free parameter whose value 
can be determined from the electroweak breaking conditions.\\
\vskip -0.3cm
        Here we discuss the minimal supersymmetric standard model
within the framework of string models, where formulae for
soft SUSY breaking parameters are given in eqs. (\ref{softm}-\ref{Aterm}) 
\footnote{Here the word ``minimal'' means only the miminal
matter content.}. In addition we take $k_3=1$, $k_2=1$ and $k_1=5/3$ for
$SU(3)$, $SU(2)$ and $U(1)$ among the standard model gauge group.
In the dilaton-dominated SUSY breaking case, i.e. $\sin \theta 
\rightarrow 1$
universal soft SUSY breaking parameters are obtained as \cite{BIM}
\begin{eqnarray}
m_i^2=m_{3/2}^2, \quad M_a=\sqrt 3 m_{3/2}, \quad
A_{ijk}=-\sqrt 3 m_{3/2}.
\end{eqnarray}
Their phenomenological implications have been studied.
On the other hand, soft SUSY breaking parameters are obtained
in the moduli-dominated SUSY breaking case, i.e.
$\cos \theta \rightarrow 1$ as
\begin{eqnarray}
m_i^2 &=& m_{3/2}^2 (1+3 \sum_m n_i^m \Theta_m^2 ), \\
M_a &=& \frac{\sqrt{3} m_{3/2}}{Re S} \sum_m (\frac{b'^m_a}{k_a}-
\delta_{GS}^m) D(T_m,T_m^*) \Theta_m, \\
A_{ijk} &=& -\sqrt{3} m_{3/2} \sum_m (1+n_i^m+n_j^m+n_k^m) \Theta_m .
\end{eqnarray}

These soft mass formulae include $n_i^m$ and $b'^m_a/k_a$,
which depend on the fields and the gauge group.
Thus these soft masses are in general non-universal.
Therefore this strong non-universality of soft masses is
an important feature of the moduli-dominated SUSY breaking.
Further, $D(T)$ works as a suppressed factor for a small value of $T$,
while soft scalar masses seem to be of $O(m_{3/2})$ naturally.   
Hence soft scalar masses are larger than gaugino masses for such
value of $T$.\\
\vskip -0.3cm
Here we have to take into account the $S-T$ mixing, that is,
at one-loop level the dilaton field and moduli fields are mixed
in the K\"ahler potential \cite{DFKZ}.
In such a case, the formulae of soft masses are obtained by 
replacing $\cos^2\theta\ \Theta_m^2$ in eq.(\ref{softm}) as \cite{BIM}  
\begin{eqnarray}
\cos^2\theta\ \Theta_m^2 \rightarrow {\cos^2 \theta\ \Theta_m^2 \over
1-a_m},
\end{eqnarray}
where $a_m=\delta_{GS}^m/24 \pi^2Y$ and $Y$ is approximately written
as $Y=S+S^*$.
This parameter $a_m$ is estimated as $a_m \approx 10^{-2}$ when
$\delta_{GS}^m=O(1)$ and $S=2$, which corresponds to 
the unified gauge coupling $\alpha_X\approx 1/25$.
Thus this parameter $a_m$ is negligible in most of the parameter space
of goldstino angles.
However, its effect is not negligible in the purely overall moduli-dominated
SUSY breaking case, which means $\cos^2 \theta =1$ and
$\Theta_m^2=1/3$ ($m=1,2,3$) exactly.
In such case the fields with the overall modular weight $\sum_m n_i^m=-1$
have the suppressed soft scalar mass as $m^2_i=am^2_{3/2}$, where we take
$a_m=a$ ($m=1,2,3$). From the viewpoint of 
multi-moduli dominated SUSY breaking the
parameter region of the goldstino angles leading to 
the suppressed soft scalar mass 
$m^2_i=am^2_{3/2}$ is rather narrow.
However, in the case of these suppressed soft masses, successful
electroweak breaking without color and/or charge breaking (CCB),
in general requires the suppressed $A$-term with the same order
of magnitude \cite{CCB,CLM}.
Such a situation, i.e. the suppressed soft scalar masses and   
the suppressed $A$-term, is effectively similar to the case with large
gaugino masses and non-suppressed soft scalar masses.
Thus we do not consider here the case of the suppressed soft scalar
masses. Instead we will discuss the effect of a large $T$ after
considering the case of $T\sim O(1)$.   
That also means effectively the case with the suppressed soft scalar 
masses.\\
\vskip -0.3cm
        Our purpose is to study implications of moduli dominated SUSY breaking,
i.e. phenomenological aspects of non-universal soft SUSY breaking
terms.
Thus we consider here the case leading to typically strong
non-universality
of soft masses.
In particular non-universality between the down and up sector Higgs
soft masses, $m_{H_1}$ and $m_{H_2}$, is interesting.
Obviously we can obtain strong non-universality between soft scalar masses
in the case that two fields have modular weights corresponding to
different moduli fields.
Hence we assume
\begin{eqnarray}
n_{H_1}=(-1,0,0), \quad n_{H_2}=(0,-1,0).
\end{eqnarray}
Even in the case with more than three moduli fields,
similar assignments can lead to maximum non-universality, e.g.
$m^2_{H_1}-m^2_{H_2} \approx m^2_{3/2}$.
Hence this case is a good example to see what happens in generic
non-universal cases derived from string models with several moduli fields.
If non-universality is not strong enough, its phenomenology is similar to
the universal case. In this case we find that
\be
m_{H_1}^2=m_{3/2}^2(1-3\sin^2\theta_1 \sin^2\theta_2),
\ee
\be
m_{H_2}^2=m_{3/2}^2(1-3\sin^2\theta_1 \cos^2\theta_2).
\ee
The inequality $m^2_{H_1} \geq m^2_{H_2}$ is favorable to realize 
successful electroweak symmetry breaking.
Thus we take here $\theta_2=0$ \ie\  $\Theta_1=0$ fixing
\begin{eqnarray}
m^2_{H_1}=m_{3/2}^2.
\l{mh1}
\end{eqnarray}
In this case we obtain the soft scalar mass of $H_2$ as
\begin{eqnarray}   
m^2_{H_2}=m_{3/2}^2(1-3\sin^2\theta_1).
\end{eqnarray}
In the case of $\sin\theta_1=0$ the universal soft mass for $m_{H_1}$
and $m_{H_2}$ is obtained, while there appears strong non-universality
around $\sin^2\theta_1 \sim 1/3$. The soft mass of $H_2$ could, in
principle, have a negative mass squared \ie\ $m^2_{H_2}<0$ with a small
magnitude at high energy scale, i.e. $\sin^2 \theta_1 \geq 1/3$ in a small
region. However, in such a case one needs a fine-tuning for other parameters.
Thus we restrict ourselves to the case of $\sin^2 \theta_1 \leq 1/3$.
As will be seen, we obtain similar results around $\sin^2 \theta \approx
1/3$. Hence we can expect similar results in the case that $\sin^2
\theta_1$ exceeds a little bit $1/3$.\\
\vskip -0.3cm 
We need modular weights of the other fields to obtain the $A$-term
and $b'^m_a$ in the gaugino mass.
For simplicity we assume that the three families of quark and lepton
fields
have $n=(-1,0,0)$.
For such a case, we are able to calculate $b'^m_a$ as
\begin{eqnarray}
b'^i_3=(-9,3,3), \quad
b'^i_2=(-8,4,5), \quad
b'^i_1=(-10,10,11).
\end{eqnarray}
Then gaugino masses are obtained as
\begin{eqnarray}
M_1= \frac{\sqrt{3} m_{3/2}}{Re S} \left[ (6-\delta_{GS})\sin\theta_1
+(33/5 -\delta_{GS}) \cos\theta_1 \right] D(T),
\nonumber\\
M_2= \frac{\sqrt{3} m_{3/2}}{Re S} \left[ (4-\delta_{GS})\sin\theta_1
+(5 -\delta_{GS}) \cos\theta_1 \right] D(T),\\
\label{gaugino1}
M_3= \frac{\sqrt{3} m_{3/2}}{Re S} \left[ (3-\delta_{GS})\sin\theta_1
+(3 -\delta_{GS}) \cos\theta_1 \right] D(T).
\nonumber
\end{eqnarray}  
In these equations we have assumed $T_m=T$ and $\delta_{GS}^m=\delta_{GS}$ 
for simplicity.
Further the $A$-term is written as
\begin{eqnarray}
A_t= -\sqrt{3} m_{3/2} \cos\theta_1.
\label{trilinear}
\end{eqnarray}
In this case soft scalar masses and $A$-terms are of $O(m_{3/2})$.   
On the other hand, gaugino masses include $D(T)$, which give a suppression
factor except a large value of $T$, i.e., $T>O(100)$.
Thus gauginos are lighter than squarks and sleptons and
the lightest neutralino and chargino are dominated by gauginos.
Which gaugino is lightest depends on $\delta_{GS}$.
It seems to be natural that the magnitude of $\delta_{GS}$
is of $O(b'^m_a)$, i.e. $O(1)$.
We have the following ratios of gaugino masses at the string scale,
\begin{eqnarray}
M_3:M_2:M_1=(3-\delta_{GS})t_1:
(1+(4-\delta_{GS})t_1):
({3 \over 5} +(6-\delta_{GS})t_1)
\end{eqnarray}
where $t_1= \tan \theta_1+1$.
When $\sin^2 \theta_1 < 1/3$, $t_1$ takes a value from 1.0 to 1.7.
At the weak scale we approximately have
\begin{eqnarray}
M_3:M_2:M_1=7(3-\delta_{GS})t_1:
2(1+(4-\delta_{GS})t_1):
({3 \over 5} +(6-\delta_{GS})t_1),
\end{eqnarray}
because $\alpha_1 : \alpha_2 : \alpha_3 \simeq 1 : 2 : 7 $ at $M_Z$.
Thus $|M_3(M_Z)|$ is larger than $|M_1(M_Z)|$ except in the case with
\begin{eqnarray}
{15 \over 6}-{1 \over 10 t_1} < \delta_{GS} <
{27 \over 8 } + {3 \over 40 t_1}.
\end{eqnarray}
Since $t_1$ is around 1.4, the above region corresponds to
$2.4 < \delta_{GS} < 3.4$.
This region is narrow.
Similarly $|M_3(M_Z)|$ is larger than $|M_2(M_Z)|$ except in the case of
$2.8 < \delta_{GS} < 3.0$.
This is also very narrow.
Therefore the gluino is heavier than the other gaugino masses except
in the very narrow region. From a phenomenological viewpoint this  
region is ruled out when $D(T)$ is not large because in this region
the gluino is the LSP.
Further $|M_2(M_Z)|$ is larger than $|M_1(M_Z)|$ unless
$2.9 < \delta_{GS} < 5.3$.
Thus we have $M_3(M_Z) > M_2(M_Z) > M_1(M_Z)$ in most parameter
space of $\delta_{GS}$ of $O(1)$, but we can obtain
$M_3(M_Z) > M_1(M_Z) > M_2(M_Z)$ in a small region.
This type of mass spectrum is derived from generic models with other
values of $b'^m_a$, that is, generally we have
$M_3(M_Z) > M_2(M_Z) > M_1(M_Z)$ in most parameter space and
$M_3(M_Z) > M_1(M_Z) > M_2(M_Z)$ in a small region,
but in a very narrow region $M_3(M_Z)$ is the smallest.
If the wino is the lightest, both of the lightest chargino and neutralino
are almost wino and these masses are degenerate \cite{CDG}.

\begin{center}
\input m1.tex
\end{center}
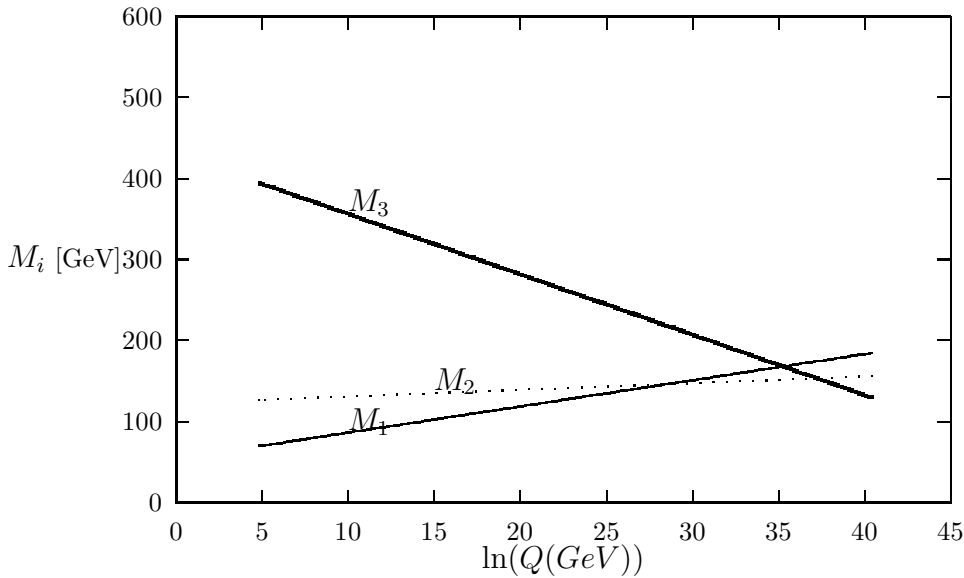
\hspace{1cm}Figure 1. The running values for $M_i$ with 
$\delta_{GS}=-5$, $m_{3/2}=500$ GeV and $\sin\theta_1=\sqrt{1 \over3}$

        In the following sections we discuss the case with
a typical value of $\delta_{GS}$ leading to   
$M_3(M_Z) > M_2(M_Z) > M_1(M_Z)$. 
Fig.1 shows the running values for $M_i$ with $m_{3/2}=500$ GeV and  
$\delta_{GS}=-5$, which is the preferred value of $\delta_{GS}$ in the 
orbifold models.\\
\vskip -0.3cm
We impose our initial conditions (\ref{mh1}-\ref{gaugino1})
at the string scale,
which differs from the gauge coupling unification scale $M_X$.
However, radiative corrections between these scales induce only small   
changes in the following discussions.
Further the difference between these two scales could be explained
by a moduli-dependent threshold correction
with a certain value of $T$ \cite{IL,stringU}.

\section{Radiative electroweak breaking and the particle
spectrum}
\hspace{0.75cm}Given the boundary conditions in eqs.
(\ref{mh1}-\ref{gaugino1}) at the compactification scale, we assume that 
$T \sim O(1)$.
Later we will comment for the cases of large $T$ of 
order 100 and small $T$ as well. Here we consider $D(T)=2.7\times 
10^{-2}$. We determine the evolution of the couplings and mass 
parameters according to 
their one loop renormalization
group equations (RGE) in order to estimate the mass spectrum of the
SUSY particles at the weak scale. The radiative electroweak
symmetry breaking scenario imposes the
following conditions on the renormalized quantities: 
\be
m_{H_1}^2 +m_{H_2}^2+2 \mu^2 > 2B \mu,
\ee
\be
(m_{H_1}^2+\mu^2)(m_{H_2}^2+\mu^2)<(B\mu)^2,
\ee
and
\be
\mu^2 = \frac{ m_{H_1}^2 -m_{H_2}^2 \tan^2\beta}{\tan^2\beta - 1}
- \frac{M_Z^2}{2},
\l{minimization1}
\ee
\be
\sin 2\beta= \frac{-2  B \mu }{m_{H_1}^2+m_{H_2}^2+ 2\mu^2 },
\l{minimization2}
\ee
where $\tan\beta= {\langle H_2^0 \rangle}/{\langle H_1^0 \rangle}$ is the
ratio of the two Higgs VEVs that give masses to the up and down type
quarks and $m_{H_1}$, $m_{H_2}$ are the two soft Higgs masses at the  
electroweak scale. It was pointed out by Gamberini et
al.~\cite{gamberini} that the tree-level effective potential $V_0$ and the
corresponding tree VEVs are strongly $Q$-dependent and the one-loop
radiative correction to $V_0$, namely
\be
\Delta V_1 = \frac{1}{64 \pi^2} \sum_{\alpha} (-1)^{2s_{\alpha}}
(2s_{\alpha}+1) C_i  M_{\alpha}^4 [ \ln 
(\frac{M_{\alpha}^2}{Q^2})-\frac{3}{2} ],
\l{deltav}
\ee
are crucial to make the potential stable against variations of the
$Q$-scale. $M_{\alpha}^2(Q)$ are the tree level mass eigenvalues,
$s_{\alpha}$ is the spin of the corresponding particle and $C_i$ is the 
color degree of freedom. \\ 
\vskip -0.3cm
        Using eqs. \refe{minimization1} and
\refe{minimization2} we can determine $\mu$ and $B$ in terms of
\mg\, $\theta_1$ and $\tan  \beta$. 
We take here $\tan \beta =2$, retaining only the top Yukawa coupling.
The value of $\vert \mu \vert $ as a function
of the gravitino mass \mg\ and the goldstino angle $\theta_1$ is given in
Fig.2. For the fixed value of $m_{3/2}$ the variation of $\vert \mu \vert
$ in this figure corresponds to different values of $\theta_1$. In the
same way, all the figures are plotted corresponding to different values of
$\theta_1$.

\vspace{0.5cm}
\begin{center}
\input mu.tex
\end{center}
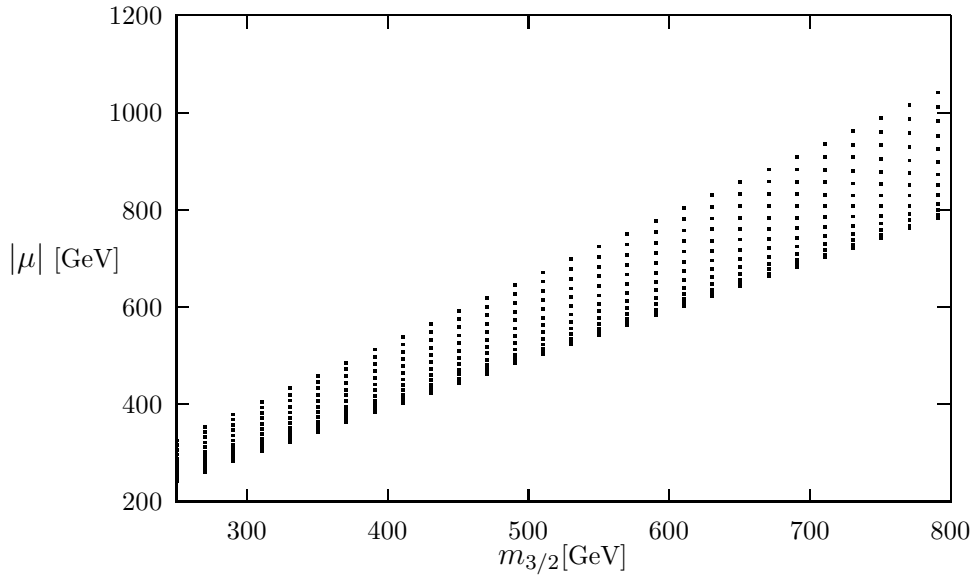
\hspace{3.5cm}Figure 2: The values of $\vert \mu \vert $ versus $m_{3/2}$ 
with  $\tan \beta=2$.
\vspace{0.5cm}

        Fig.3 shows the ratio of the coefficient of the bilinear 
term $B$ at the
compactification scale to \mg\ versus the gravitino mass. We note that
the sign of $B$ in general is opposite to that of $\mu$ 
for the realization of electroweak symmetry breaking. 
It is also remarkable that the value of 
$B/m_{3/2}$ is very stable against $m_{3/2}$, \ie\ $B/m_{3/2}\sim 0.35 
(-1.6)$ for $\mu >0 (\mu <0)$. That could suggest the effectiveness of 
a certain type of $\mu$-term generation mechanism.

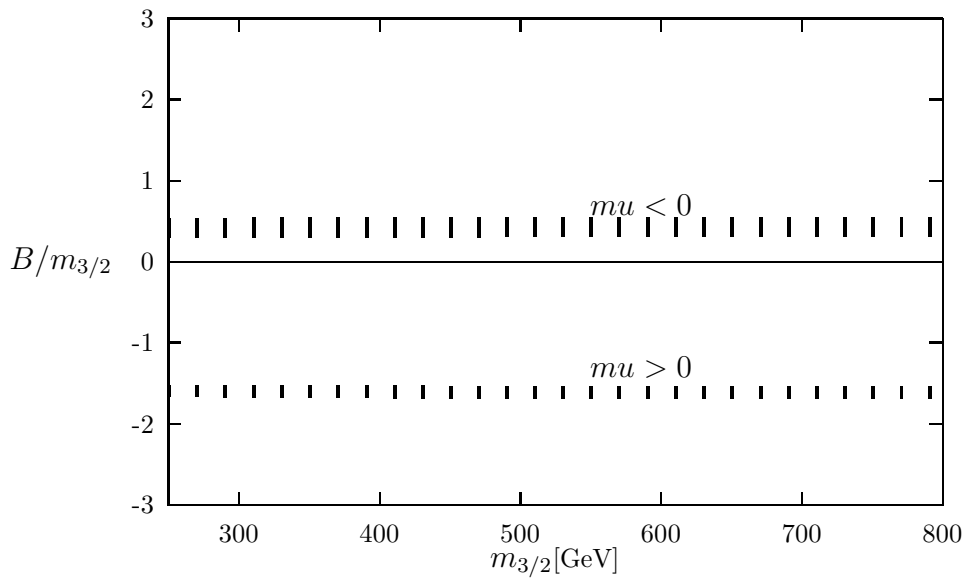
\begin{center} 
\input b.tex
\end{center}
\hspace{3.5cm}Figure 3: The values of $B/m_{3/2} $ versus $m_{3/2}$ with  
$\tan \beta=2$.
\vspace{0.5cm}
 
        An important constraint on the parameter space arises from the 
experimental lower bound on the chargino mass from \LEP2 , 
$m_{\chi^{\pm}}   
>84$ GeV. This bound is applied as long as $m_{\chi^{\pm}}-m_{\chi} > 3$ 
GeV which is always satisfied in this model. Here $m_{\chi}$ is a
neutralino mass. As we can see from 
Figs.4 and 5 this bound implies that the gravitino mass has to
be $m_{3/2}>280$ GeV for $\mu >0$ and $ m_{3/2}>420$ GeV for $\mu <0$.
\vspace{0.5cm} 
\begin{center}
\input w1.tex
\end{center}
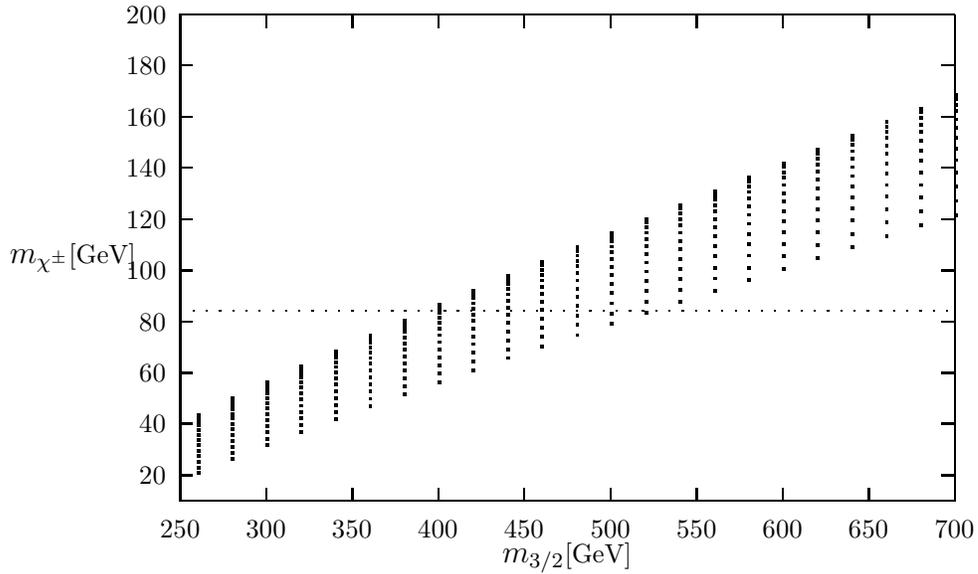
\hspace{3.5cm}Figure 4: The lightest chargino mass versus $m_{3/2}$ with
$\mu<0$.
\vspace{0.5cm}
\vspace{0.5cm} 
\begin{center}
\input w2.tex
\end{center}
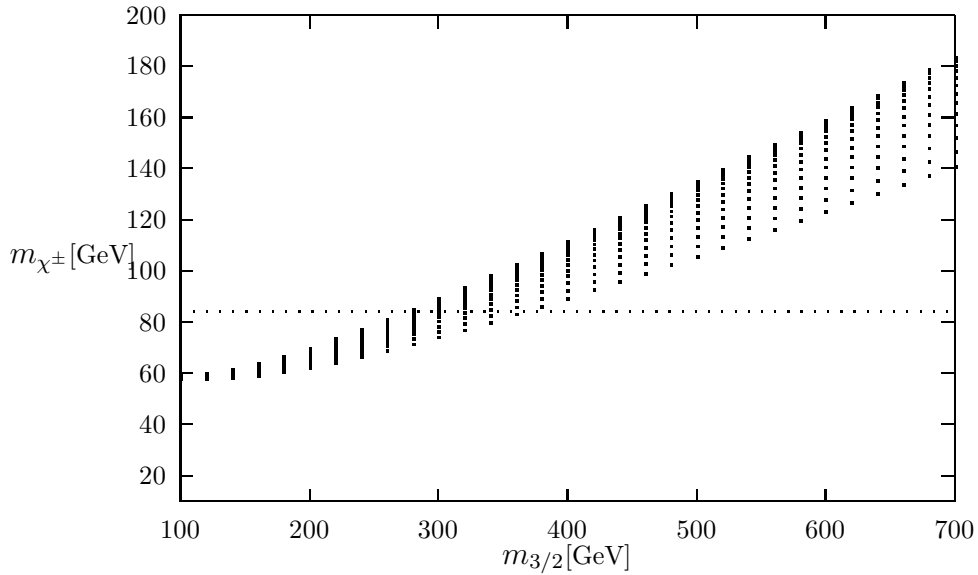
\hspace{3cm}Figure 5: The lightest chargino mass of versus 
$m_{3/2}$ with $\mu>0$.
\vspace{0.5cm}

It is clear that at these values of the
gravitino mass  most of the scalar particles become very heavy. For
instance, the right selectron mass is of order 300 GeV at this lower bound
of the 
gravitino mass in this model, while the right selectron was found to be 
the lightest sfermion in the case of dilaton contribution to SUSY 
breaking~\cite{Khalil}. Moreover, the off diagonal element of the stop 
mass matrix  
$m_t(A_t+\mu \cot \beta)$ is comparable to the diagonal parts of this
matrix. This gives a chance to have one of the stops to be light and   
Fig.6 shows the values of this stop mass versus the gravitino mass
for $\mu >0$ since for this case we have maximum mixing.
\vspace{0.5cm} 
\begin{center}
\input mt1.tex
\end{center}
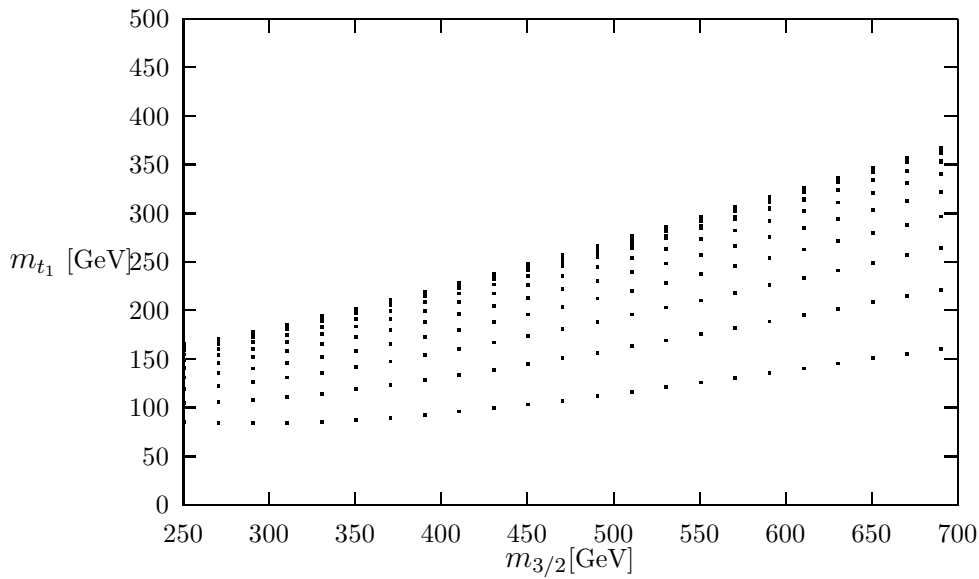
\hspace{3cm}Figure 6: The mass of the lightest stop quark  versus 
$m_{3/2}$ with $\mu >0$.
\vspace{0.5cm}

Actually this light stop is predicted to be the lightest sfermion 
in this class
of models and as we will show below it has important effects on the    
phenomenological implication such as the prediction of the branching
ratio of the $b\rightarrow s \gamma $ decay and the relic abundance of  
the LSP. We would like to stress that this feature, having a
significant splitting between the masses of the stop quarks, is absent in
the overall modulus SUSY breaking scenario and dilaton contribution to
SUSY breaking scenario. \\
\vskip -0.3cm 
        Now we would like to investigate the composition and the mass of
the lightest neutralino. The lightest neutralino $\chi$ is a linear 
combination of two neutral gauginos $\tilde{B}^0$ (bino) and $\tilde{W}_3^0$ 
(wino) and the Higgsinos $\tilde{H}^0_1$, $\tilde{H}^0_2$,
$$ \chi= N_{11} \tilde{B}^0 + N_{12} \tilde{W}_3^0 + N_{13} \tilde{H}^0_1 + 
N_{14} \tilde{H}^0_2,$$ where $N_{ij}$ are the entries of the unitary
matrix 
which diagonalizes the neutralino mass matrix and they are functions of 
$\tan\beta$ , $M_2$ and $\mu$. The gaugino `purity' function $f_g=\vert 
N_{11}\vert^2 + \vert N_{12} \vert^2 $ describes the neutralino 
composition. As explained, in most 
of the parameter space the wino is heavier than the bino. Thus we find
that the lightest neutralino is mostly bino \eg, $f_g=0.99$ for
$\delta_{GS}=-5$.\\
\vskip -0.3cm
The lightest neutralino mass  corresponding to the gravitino mass     
is plotted in Figs.7 and 8 for $\mu > 0$ and $\mu <0$   
respectively. From these figures we can easily realize that the
lightest neutralino is indeed the LSP and this has an important 
cosmological implication as we will see in section 5. 
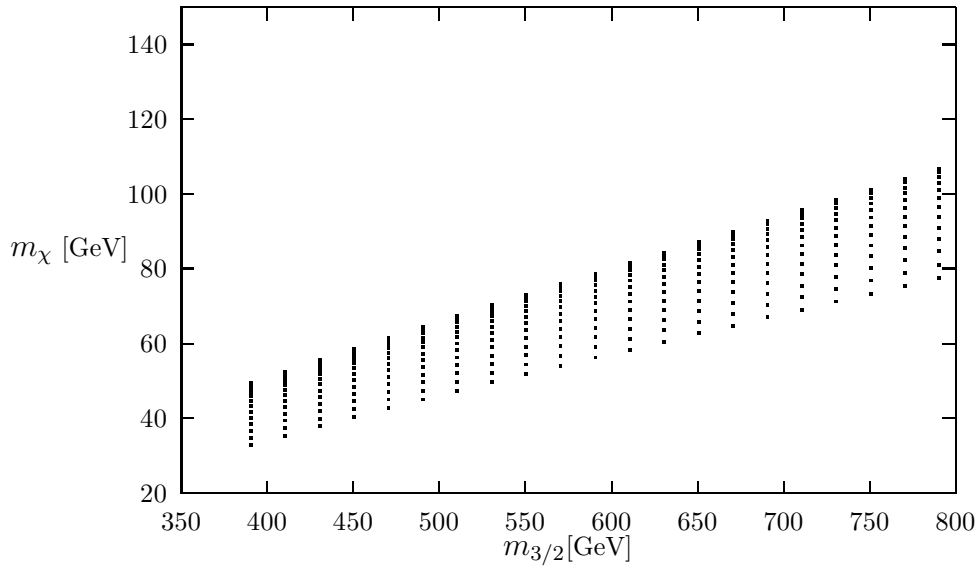
\vspace{0.5cm}
\begin{center}
\input lsp1.tex
\end{center}
\hspace{2cm}Figure 7: The lightest neutralino as a function of the 
gravitino mass for $\mu >0$.
\vspace{0.5cm}
\begin{center}
\input lsp2.tex
\end{center}
\hspace{2cm}Figure 8: The lightest neutralino as a function of the
gravitino mass for $\mu <0$.
\vspace{0.5cm}

 Also we are interested in the SUSY Higgs spectrum, and
in particular the lightest Higgs scalar $h$ and charged Higgs
scalar $H^+$ whose mass is very relevant for $b\rightarrow s
\gamma$ branching ratio. The lightest Higgs mass is given
by\cite{haber}
\be
m_h^2=m_{h^0}^2 + (\Delta m_h^2)_{1LL} + (\Delta m_h^2)_{mix},
\l{1lop}
\ee   
\be
m^2_{h^0}= \frac{1}{2} \left( m_A^2+m_Z^2 - \sqrt{(m_A^2+m_Z^2)^2- 4 m_Z^2
m_A^2 \cos^2 2\beta} \right),
\ee
where $m_A^2=m^2_{H_1}+m^2_{H_2}+2 \mu^2$,
\be
(\Delta m_h^2)_{1LL}= \frac{3 m_t^4}{4\pi^2 v^2}
\ln(\frac{m_{\tilde{t}_1}m_{\tilde{t}_2}}{m_t^2})
\left[1+O(\frac{m_W^2}{m_t^2})\right],
\ee
\be     
(\Delta m_h^2)_{mix}= \frac{3 m_t^4 \tilde{A}_t^2}{8\pi^2 v^2}
\left[ 2h(m_{\tilde{t}_1}^2,m_{\tilde{t}_2}^2)+ \tilde{A}_t^2
f(m_{\tilde{t}_1}^2,m_{\tilde{t}_2}^2)\right]
\left[1+O(\frac{m_W^2}{m_t^2})
\right],
\ee
with $\tilde{A}_t=A_t +\mu \cot \beta$. 
The functions $h$ and $f$ are given by \be
h(a,b) = \frac{1}{a-b} \ln(\frac{a}{b}) \hspace{1cm} and
\hspace{0.2cm}
f(a,b)=\frac{1}{(a-b)^2}\left[2-\frac{a+b}{a-b}
\ln(\frac{a}{b})\right].
\ee
The two loop leading logarithmic contributions to $m_h^2$ are 
incorporated by
replacing $m_t$ in eq.\refe{1lop} by the running top quark mass
evaluated at the scale $\mu_t$ which is given by $\mu_t =\sqrt{M_t
M_s}$ where $M_t$ is the pole mass of the top quark, $M_t=174$ GeV, and
$M_s=\sqrt{\frac{M_{\tilde{t_1}}^2+M_{\tilde{t_2}}^2}{2}}$.
Fig.9 shows the lightest Higgs mass as a function of the
gravitino mass with $\mu >0$ where we have a maximum mixing.  

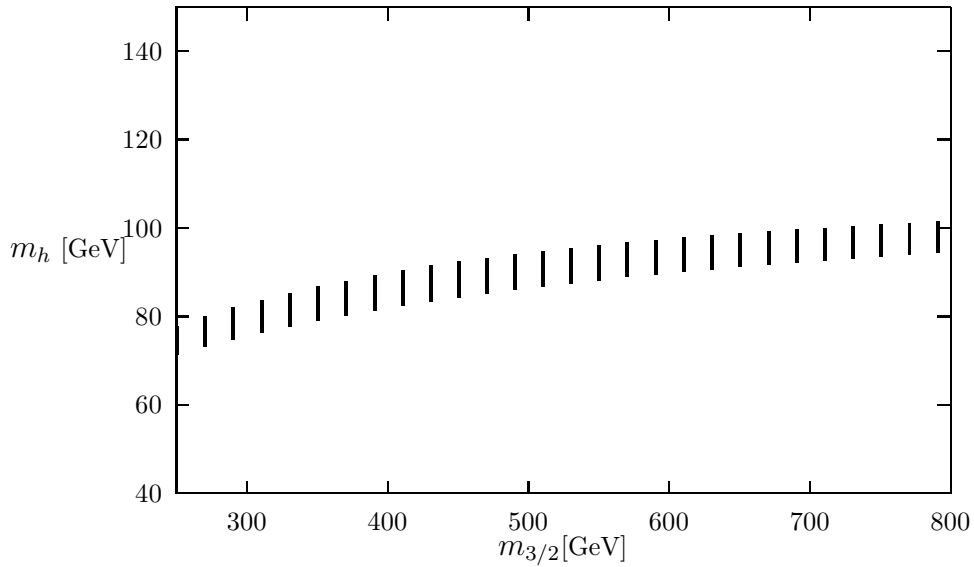
\begin{center}
\input mhiggs.tex
\end{center}
\hspace{1.5cm}Figure 9: The lightest Higgs mass as a function of the 
gravitino mass.
\vspace{0.5cm}

It shows that the lower bound of this mass is of
the order 70 GeV. The charged Higgs mass is given by
$$m_{H^{\pm}}^2 =m_W^2+m_{H_1}^2+m_{H_2}^2 +2 \mu^2.$$
Fig.10 gives the charged Higgs mass versus the gravitino mass.
\vspace{0.5cm}
\begin{center}
\input chg.tex
\end{center}
\hspace{2cm} Figure 10: The charged Higgs mass versus the gravitino mass.
\vspace{0.5cm}

It is interesting to note that in this class of models the charged Higgs 
field is
very heavy. This is due to the fact that $m_{H_1}^2$ is much larger
(and positive) than $m_{H_2}^2$ (which is negative at the weak scale),
and $\mu$ is quite large. 
As we will see, the evidence of having heavy charged Higgs field 
gives an interesting implication in studying the constraints on the
parameter space due to the $b \rightarrow s \gamma$. Also we have to
stress that if $m_{H_1}^2$ is of the order $m_{H_2}^2$ as in the case of
the overall modulus and dilaton SUSY breaking scenarios, we get a
cancellation
between them and the charged Higgs field becomes much lighter. So we can
conclude that the strong non-universality between the Higgs soft masses at
GUT scale is preferable.
\section{Constraints from $b \rightarrow s \gamma$}
\hspace{0.75cm}In this section we focus on the constraints on the
parameter space (\mg\ , $\theta_1$) which arise from the $b \rightarrow s
\gamma$ decay since the CLEO observation \cite{amer} confirmed that $
1\times 10^{-4} < BR(b \rightarrow s \gamma) < 4 \times 10^{-4}$. It is
well known that in supersymmetric models there are three significant
contributions of total amplitude from
the $W$-loop, charged Higgs loop and chargino loop. The
inclusive branching ratio for $b \rightarrow s \gamma$ is given by
\be
R = \frac{BR(b \rightarrow s \gamma)}{BR(b \rightarrow c e\bar{\nu})}.
\ee
The computation of $R$ yields~\cite{bs}
\be
R= \frac{\mid V_{ts}^* V_{tb}\mid^2}{\mid V_{cb} \mid^2} \frac{6
\alpha_{em}}{\pi} \frac{[\eta^{16/23} A_{\gamma} + \frac{8}{3} (\eta^{14/23}
-\eta^{16/23}) A_g + C]^2}{I(x_{cb})[1-\frac{2}{3\pi} \alpha_S(m_b)
f(x_{cb})]}.
\ee
Here, $\eta$ is the ratio of the running strong-coupling constants at 
two energy scales $m_W$ and $m_b$ \ie \  $\eta =
\frac{\alpha_S(m_W)}{\alpha_S(m_b)}$. $C$ represents the leading-order QCD
corrections to $b \rightarrow s \gamma$ amplitude at the $Q=m_b$
scale~\cite{misiak2}. The function $I(x)$ is given by
$$I(x)=1-8 x^2 +8 x^6 -x^8 -24 x^4 \ln x, $$
and $x_{cb} = \frac{m_c}{m_b}$, while $f(x)$ is a QCD correction
factor $ f(x_{cb})=2.41$. The amplitude $A_{\gamma}$ is from
the photon penguin vertex, the amplitude $A_g$ is from the gluon penguin
vertex and they are given in Ref.~\cite{bs}. It was shown that in
MSSM~\cite{bs}, pure dilaton SUSY breaking~\cite{stringbs}   
and minimal string unification~\cite{Khalil} with $\tan \beta$ of
order 2 the chargino contribution gives rise to a destructive interference 
(in case of $\mu <0$ ) with SM contribution and charged Higgs contribution
but it
is generally smaller than the charged Higgs contribution.
This leads to
a severe constraint on the parameter space of these models. Also it was
realized
that the constraint is less severe in the case of
the non-universality between the soft terms than that for the universal
case.
However, in minimal string unification the non-universality is very tiny 
since the allowed values of goldstino angle \th\ is very close to 
$\pi/2$ which
corresponds to purely dilaton-dominated SUSY breaking, namely $\theta \in 
[0.98 rad. , 2 rad.]$.   
This constraint on \th\ arised from the avoidance of the tachionic mass at
string scale and the conservation of the electric charge. \\
\vskip -0.3cm
        Before we present the result of $b \rightarrow s \gamma$ in 
the three moduli dominated SUSY breaking, we find that it is 
worthwhile to study first this constraint on the case of the overall modulus
since it was omitted in Ref.~\cite{CDG}. To be able to compare the result of 
the overall modulus with that of $b \rightarrow s \gamma$  in our case 
where $\delta_{GS}=-5$ we have to use the corresponding value of 
$\delta_{GS}$ in the overall modulus scenario which is $\delta_{GS}=-15$. 
We find that in this case the chargino mass reaches $84$ GeV at very 
large values of the gravitino mass \mg: about 400 GeV for $\mu >0$ 
and 700 Gev for $\mu<0$. Figs.11 and 12 show the branching ratio of 
$b\rightarrow s \gamma$ versus the gravitino mass in the case of the
overall modulus dominated SUSY breaking for $\mu >0$ and $\mu <0$
respectively.
\vspace{0.5cm}
\begin{center}
\input chen1.tex
\end{center}
\hspace{1.5cm}Figure 11: The branching ration of $b \rightarrow s \gamma$ 
as a function of \mg\ for $\mu>0$.
\vspace{0.5cm}
\begin{center}
\input chen2.tex
\end{center}
\hspace{1.5cm}Figure 12: The branching ration of $b \rightarrow s \gamma$   
as function of \mg\ for $\mu<0$.  
\vspace{0.5cm}

        From these figures it is evident that the CLEO upper bound imposes  
severe constraints on the allowed parameter space of the overall modulus 
dominated SUSY breaking. This behaviour is independent of the choice of
the  value of $\delta_{GS}$. In the case of $\mu>0$ the
value of 
the branching ratio of $b  \rightarrow s \gamma$ falls outside the 
experimentally 
allowed region for all the values of \mg\ up to $1$ TeV. On the other 
hand, for $\mu 
<0$, where we have $m_{3/2}>700$ GeV, we find that the branching ratio of 
$b \rightarrow s \gamma$ becomes less than the CLEO upper bound at 
$m_{3/2} >1.4$ 
TeV. It is clear that at this value of \mg\ the charged Higgs field 
becomes very heavy and its contribution becomes small. \\
\vskip -0.3cm

        Now we turn to our model. We find that the chargino contribution 
gives rise to substantial destructive interference with SM and $H^+$ 
amplitude. At $\tan \beta=2$  Figs.13 and 14 show the $b\rightarrow s 
\gamma$ branching ratio for $\mu >0$ and $\mu<0$ respectively. 
\vspace{0.5cm}
\begin{center}
\input mbs1.tex
\end{center}
\hspace{1.5cm} Figure 13: The branching ratios of $b \rightarrow s 
\gamma$ versus \mg\ and $\mu >0$.

\vspace{0.5cm}
\begin{center}
\input mbs2.tex
\end{center}
\hspace{1.5cm} Figure 14: The branching ratios of $b \rightarrow s
\gamma$ versus \mg\ and $\mu <0$.
\vspace{0.5cm} 

 From these figures we find that
the branching ratios of $b \rightarrow s \gamma$ in this model for $\mu 
>0$ are less than the SM value and we conclude that there is no   
essential constraint from $b \rightarrow s \gamma$ imposed on the      
parameter space. We would like to emphasize the reasons of the wining of
the
chargino contribution. First, as we have mentioned and as Fig.10 shows,
the mass of the charged Higgs field is quite heavy because of the strong
non-universality between Higgs masses $m_{H_1}^2$ and $m_{H_2}^2$. For 
instance, if the chargino mass is equal to 84 GeV, the charged Higgs mass 
is around 400 GeV, so that the charged Higgs contribution which
is inversely proportional to its mass square becomes quite small.
Second, in this model we have a significant
splitting between the values of the stop masses, as we mentioned due  
to the large mixing in the stop mass matrix. In fact, the chargino 
amplitude crucially depends on this splitting.\\
\vskip -0.3cm
        This result is quite interesting since, as it was pointed
out in~\cite{misiak}, the SM prediction is above the CLEO measurement at
the $1 \sigma$ level. The physics beyond the SM should provide a
destructive interference with the SM amplitude and 
our model has this feature with $\mu >0$.
\section{Relic abundance of the lightest neutralino}
\hspace{0.75cm}We have shown that the lightest neutralino turns out to be
the LSP and it is mostly pure bino. So it could provide a natural source
for the dark matter required by galactic rotational    
data. In this section we would like to study the relic abundance of the
LSP and investigate the constraints on the parameter
space by requiring the neutralino relic density to be $0.1 \leq
\Omega_{LSP} \leq 0.9$ with $0.4 \leq h \leq 0.8$. It was shown in
Ref.~\cite{Khalil} that these values of $\Omega_{LSP}$ impose a stringent 
upper bound on the parameter space in the minimal string unification
namely it leads to an upper bound on the gravitino mass of about 600
GeV.\\
\vskip -0.3cm
        Since the LSP is mostly pure bino, 
its coupling to the lightest Higgs field   
and $Z$ boson is weak and the sfermions are very heavy, 
many channels of the neutralino annihilation are closed or suppressed.  
We find that the annihilation process is dominated by the exchange of the
lightest stop into up type quarks \ie\ only $u$ and $c$. The 
$t$-channel can not be opened because the neutralino mass is smaller 
than the mass of the top quark. As we explained the stop is the lightest
sfermion for this model. 
The $Z$-boson contribution is suppressed, except for
$m_{\chi} \sim m_Z/2$ due to the small $Z \chi \chi$ coupling  
$ (ig/2 \cos\theta_W) (N_{13}^2-N_{14}^2) \gamma^{\mu} \gamma_5$.\\
\vskip -0.3cm
        In the overall modulus dominated SUSY breaking, both stop 
masses are quite large. The $\chi \chi $ annihilation is very
small and this leads to a very large relic density of order
$10^2$ \cite{CDG}, which is, of course, an unacceptable value. In
Ref.\cite{CDG} the co-annihilation between the LSP and the chargino was
considered to reduce these values of the relic density. However, Fig.17
in this reference shows that for $\tan\beta=2$ a small part of the
parameter space can lead to $\Omega_{LSP} h^2 \leq 1$, moreover this part
corresponds to chargino mass less than 88 GeV.\\ 
\vskip -0.3cm
  
        For the computation of the lightest neutralino relic abundance  we
have to determine the  thermally averaged cross section $\langle \sigma_A
v \rangle \sim a + b v$ \cite{grist}. By neglecting the
fermion masses with respect to the LSP mass we find that 
$a=0$, and $b$ is given in Ref.\cite{grist}. Given $a$ and $b$ we can
determine the relic LSP density
$\Omega_{\chi} h^2$:
\be
\Omega_{\chi} h^2 = \frac{\rho_{\chi}}{\rho_c/h^2} = 2.82 \times 10^8
Y_{\infty} (m_{\chi}/GeV),
\ee
where
\be
Y_{\infty}^{-1}=0.264\ g_*^{1/2}\ M_P\
m_{\chi}\ (\frac{a}{x_F}+\frac{3b}{x_F^2}),
\ee
and $h$ is the Hubble parameter.
We take  $ 0.4 \leq h \leq 0.8$,
and $ \rho_c \sim 2 \times 10^{-29} h^2$ is the critical density of the
universe. In addition the freeze-out epoch $ x_F$ is written as $x_F= 
m_{\chi}/T_F$ where $T_F$ is the freeze-out temperature
and the $\chi \chi $ annihilation rate is smaller than the
expansion rate of the universe below $T_F$. 
We can iteratively compute the freeze-out
temperature from
\be
x_F= \ln \frac{0.0764 M_P ( a+ 6 b/ x_F) c ( 2+c) m_{\chi}}{\sqrt{g_*
x_F}}.
\ee
Here $ M_P= 1.22 \times 10^{19} $ GeV is the Planck mass and $ g_*$ ($ 8
\leq \sqrt{g_*} \leq 10$) is the effective number of
relativistic degrees of freedom at $T_F$.
\begin{center}
\input relic.tex
\end{center}
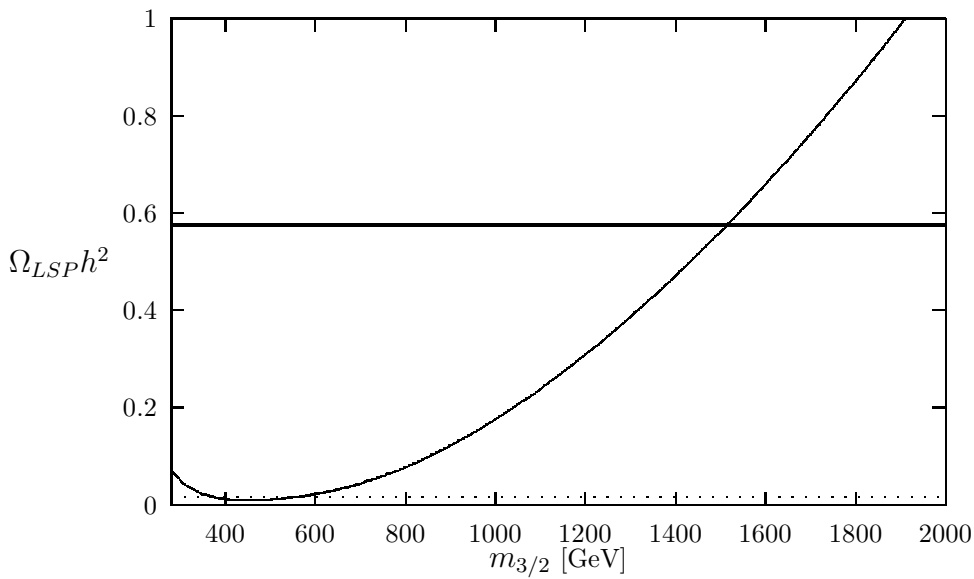
 Figure 15: The neutralino relic abundance versus \mg\ 
where $\tan\beta=2$, $\sin\theta_1\sim \sqrt{1\over 3}$ and $\mu >0$. The
solid line corresponds to the maximum value of $\Omega_{\chi}h^2$ we
assumed and the dotted line corresponds to the minimum value.
\vspace{0.5cm}
 
Fig. 15 shows the relic abundance of the lightest neutralino
$\Omega_{\chi}h^2$ as a function of the gravitino mass where the goldstino
angle $\theta_1$ is equal to $\sin^{-1}\sqrt{1\over 3} \sim 0.6\ rad$. We
find that the requirement of the neutralino relic density to be $ 0.1 \leq
\Omega_{LSP} \leq 0.9$ and $ 0.4\leq h\leq 0.8$ imposes an upper bound 
1.5 GeV on the gravitino mass. Also we find that the
neutralino relic density imposes a severe constraint on $\theta_1$, namely
$\theta_1 \geq 0.4$ rad. For smaller values of $\theta_1$ we obtain
$\Omega_{\chi}h^2>1$. This means that the neutralino relic abundance of
the moduli SUSY breaking does not prefer the universality between 
$m_{H_1}$ and $m_{H_2}$.   
\section{Effects of other values of $T$ and $n_k^i$}
\hspace{0.75cm}  In this section we analyze effects of other values of        
$T$ and $n_k^i$, because we have left them aside in the
previous sections.
They could lead to an interesting implication or there
could be some constraints on these values.
Obviously we can obtain similar phenomenological results in the case
of $T$ of $O(1)$ except the very small $T$ case like $T=1.2$ and
$D(T)=1.5 \times 10^{-3}$.
Similar results are obtained even in the case with $T=10$, 20 or some
small value of $O(10)$.
Furthermore, the results in the previous sections do not depend on a
value of $\delta_{GS} \sim O(1)$ sensitively.
We obtain almost the same results for other values of $\delta_{GS}$
leading to $M_3(M_Z)>M_2(M_Z)>M_1(M_Z)$. \\
 
\hspace{-0.75cm}{\large{\bf The case of small $T$:}} Gaugino masses become
much smaller compared with sfermion masses and  $A$-term as well as the
gravitino mass. The lightest chargino and neutralino become almost purely
gauginos.
So the experimental lower bound of the chargino mass pushes
the gravitino mass to a higher value.
For example, the case of $T=1.2$ requires the gravitino mass above 4
TeV. Then the mass of the
scalar particles are very heavy and this is an undesirable feature for
the hierarchy
problem and naturalness.
Moreover in this case most of the lightest neutralino
annihilation channels are suppressed or closed so that the relic
abundance $\Omega_{LSP} h^2$ is very large.\\
 
\hspace{-0.75cm}{\large{\bf The case of large $T$:}} Gaugino masses become
comparable to sfermion masses and
$A$-term as well as the gravitino mass, e.g. in the case
of $T \sim O(100)$.
Note that this value of $T$ has the effective meaning including the
case of $T \sim O(1)$ and the suppressed $m_i$ due to the $S-T$ mixing
as said in section 2.
The phenomenological prediction of
this model becomes rather similar to that of the universal
model. The gaugino purity function becomes smaller.
In this case the experimental lower bound of the chargino mass
does not always require a larger gravitino mass, that is,
$O(100)$ GeV is enough.
RGE effects due to gaugino masses become important.
For example, non-universality of scalar masses at the string scale  
is diluted at the weak scale because of RGE effects of large 
gaugino masses. If we get very small non-universality between two Higgs
masses,
fine-tuning should be required to realize successful electroweak
symmetry breaking.
For this type of mass spectrum, the charged Higgs field also contributes to
the branching ratio of $b \rightarrow s \gamma$.
In such a case we can not expect the same results as those in 
the case with $T=5$.\\

\hspace{-0.75cm}{\large{\bf Other values of $n_k^i$:}} We have taken
$n=(-1,0,0)$ for all the matter fields in the 
analyses of the previous sections.
In a similar way we can analyze the case where matter fields have 
other modular weights.
In such case, the parameter space of the gravitino mass and the 
goldstino angles can have further constraints, e.g. 
experimental lower bounds of sfermion masses, degeneracy of 
sfermion masses for flavor changing processes and the 
constraint to avoid $m_i^2 <0$ or CCB.\\
\vskip -0.3cm
In addition, the anomaly coefficients $b'_a$ are changed when we 
alter the assignment of modular weights.
However, the structure of the gaugino mass spectrum, 
i.e. $M_3(M_Z) > M_2(M_Z) >M_1(M_Z)$, is not sensitive to 
$b'_a$ of $O(1)$.
Thus we can obtain similar results for the gaugino-Higgs sector 
in the case with other assignments of modular weights.
  
\section{Conclusions}

We have studied phenomenological implications of moduli-dominated
SUSY breaking.
In general, moduli-dominated SUSY breaking leads to non-universal soft
scalar masses as well as non-universal gaugino masses.
In addition, gauginos are lighter than most sfermions.
This type of mass spectrum is very different from the one derived
from dilaton-dominated SUSY breaking and the mixed dilaton/moduli
breaking
case as well as ordinary ``minimal'' supergravity with universal soft
breaking terms.
Thus moduli-dominated SUSY breaking leads to phenomenological aspects
different from other types of SUSY breaking.\\
\vskip -0.3cm
Non-universality between two Higgs masses is favorable to realize successful
radiative electroweak symmetry breaking.
In moduli-dominated SUSY breaking the lightest chargino and neutralino
are almost gauginos and the latter is usually regarded as the LSP.
Further there appears a mass splitting between the lightest Higgs field
and the other Higgs fields, e.g., the charged Higgs field.
Also one of the stop fields is very light compared with the other stop
as well as other squarks and sleptons.
This type of mass spectrum makes the branching ratio of
$b \rightarrow s \gamma$ decrease, while the overall moduli case
is ruled out in a wide parameter space.
Furthermore, strong non-universality between $m_{H_1}$ and $m_{H_2}$
is favorable for the constraint from the relic abundance of the LSP.  
Hence moduli-dominated SUSY breaking is the very interesting
case in the whole parameter space of goldstino angles of string models.\\
\vskip -0.3cm
When we take a very small value of $T$, a mass splitting
between gauginos and others becomes large.
Thus the lower bound of the chargino mass requires large sfermion masses.
In this case the cosmological constraint from the relic abundance of
the LSP plays a role to rule out this parameter space.
On the other hand, gaugino masses are comparable to sfermion masses 
in the case of $T \sim O(100)$.
In this case large gaugino masses dilute non-universality among
soft scalar masses because of RGE effects.
Thus we obtain phenomenological aspects similar to the universal case.
In such case the charged Higgs field also contributes to the
$b \rightarrow s \gamma$ decay increasing its branching ratio.\\
\vskip -0.3cm 
If gauge symmetries break reducing their ranks, there appears
another type of contribution to soft scalar masses, i.e.
$D$-term contributions \cite{Dterm1,Dterm2}.
These $D$-contributions can also become sources of non-universality
among soft scalar masses.
Thus analyses including these $D$-term contributions would be
interesting \cite{GUTD,stringD}.\\

\noindent{\Large\bf Acknowledgements}
\vskip0.5truecm
 The authors are grateful to D.Suematsu and Y.Yamagishi for useful
discussions. S.K would like to acknowledge the hospitality of ICTP.

\end{document}